\pdfoutput=1

\documentclass[12pt]{article}
\usepackage{acronym}

\usepackage{a4wide}
\usepackage{scalefnt}
\usepackage{amsmath}
\usepackage{amssymb}
\usepackage{dsfont}
\usepackage[labelformat=simple]{subcaption}
\usepackage{graphicx}
\usepackage{listings}
\usepackage{xspace}
\usepackage{url}
\usepackage{hyperref}
\usepackage{fancyhdr}
\usepackage{booktabs}
\usepackage{longtable}
\usepackage[numbers,sort&compress]{natbib}
\usepackage{filemod}
\usepackage[affil-it]{authblk} \usepackage[usenames,dvipsnames]{color}
\usepackage[final]{showlabels}
\usepackage[utf8x]{inputenc}
\showlabels{cite}

\usepackage[usenames,dvipsnames]{color}
\fancypagestyle{firstpage}
{
  
  \fancyhead[R]{\myheaderline}
}
\newcounter{notecount}
\newcommand{\mbottom}{m_\mathrm{b}}
\newcommand{\citere}[1]{Ref.\,\cite{#1}}
\newcommand{\citeres}[1]{Refs.\,\cite{#1}}
\newcommand{\pt}{p_\perp}
\newcommand{\one}{one}
\newcommand{\two}{two}
\newcommand{\three}{three}
\newcommand{\note}[2]{ {\stepcounter{notecount}\sf%
    \footnotesize\color{red}[\arabic{notecount}$|${\color{blue}#1}$|$#2]}}
\renewcommand{\note}[2]{}
\acrodef{QCD}{Quantum Chromo Dynamics}
\acused{QCD} 
\acrodef{IBP}{integration-by-parts}
\acrodef{BSM}{beyond-the-\ac{SM}}
\acrodef{LHC}{Large Hadron Collider}
\acrodef{LO}{leading order}
\acrodef{NLO}{next-to-leading order}
\acrodef{NNLO}{next-to-next-to-leading order}
\acrodef{PDF}{parton density function}
\acrodef{SM}{Standard Model}
\newcommand{\mtop}{m_\mathrm{t}}
\newcommand{\mquark}{m_\mathrm{q}}
\newcommand{\mhiggs}{M_\mathrm{H}}
\newcommand{\myheaderline}{%
  \small\sf
  July 2019\\
  TTK-19-26\\
  P3H-19-021\\
  FR-PHENO-2019-011\\
}

\makeatletter
\DeclareRobustCommand*{\bfseries}{%
   \not@math@alphabet\bfseries\mathbf
   \fontseries\bfdefault\selectfont
   \boldmath
}
\makeatother

\numberwithin{equation}{section}

\newcommand{\order}[1]{\ensuremath{\mathcal{O}(#1)}}
\newcommand{\ggh}{\ensuremath{ggH}\xspace}
\newcommand{\aah}{\ensuremath{\gamma\gamma H}\xspace}
\newcommand{\MSbar}{\ensuremath{\overline{\text{MS}}}\xspace}

\title{The light-fermion contribution to the exact Higgs-gluon form factor in
QCD }

\author[1]{Robert V. Harlander}
\author[2]{Mario Prausa}
\author[3]{Johann Usovitsch}
\affil[1]{Institute for Theoretical Particle Physics and
  Cosmology,\protect\\ RWTH Aachen University, D-52056 Aachen, Germany}
\affil[2]{Physikalisches Institut,
Albert-Ludwigs-Universit\"at,\protect\\
D-79085 Freiburg, Germany}
\affil[3]{Trinity College Dublin, School of Mathematics, Dublin 2, Ireland}

\date{}

\begin{document}
\maketitle
\thispagestyle{firstpage}
\begin{abstract}
  An analytical expression for the \three-loop form factors for $\ggh$
  and $\aah$ is derived for the contributions which involve massless
  quark loops. The result is expressed in terms of harmonic
  polylogarithms. It fully agrees with previously obtained kinematical
  expansions, and confirms a recent semi-numerical approximation which
  extends over the full kinematic range.
\end{abstract}



\section{Introduction}

The study of the Higgs boson is one of the most promising ways to search
for physics beyond the \ac{SM}.  A necessary precondition for this to be
successful is the precise understanding of the relevant \ac{SM}
predictions. One of the most important quantities in this respect is the
cross section for Higgs production in gluon fusion. In fact, significant
theoretical efforts have been made to pin down its \ac{SM} value, and to
estimate the associated uncertainties (see \citere{Cepeda:2019klc} for a
recent review). One source of uncertainties is the fact that, up to now,
\ac{QCD} corrections to the Higgs cross section beyond \ac{NLO} are
based on the approximation of an infinitely heavy quark mediating the
gluon-Higgs coupling. For the top-quark contribution, which by far
dominates the total cross section at the \ac{LHC}, comparison of this
limit to the full result at \ac{NLO} shows agreement at the sub-\% level
for a Higgs mass of $\mhiggs=125$\,GeV, providing confidence in using
this approximation also at higher orders of perturbation
theory~\cite{Spira:1997dg,Spira:1995rr}. In fact, an explicit
calculation of sub-leading terms in $1/\mtop$ at \ac{NNLO}, combined
with the high-energy limit of the cross section, further justifies this
procedure~\cite{Harlander:2009my,Harlander:2009mq,Pak:2009dg,Marzani:2008az}.
Nevertheless, the lack of the exact top mass dependence still requires
one to associate with it an uncertainty on the total cross section of
the order of 1\%. It is thus a non-negligible contribution to the
overall uncertainty of about 5\%, which also includes uncertainties
induced by \acp{PDF} and $\alpha_s$, for example (see
\citeres{deFlorian:2016spz,Anastasiou:2016cez}).

A related uncertainty arises from the bottom-quark induced Higgs-gluon
coupling. While suppressed by the bottom Yukawa coupling, its effect on
the \ac{LO} cross section is still a reduction by about 6\%. Since the
numerical value of the bottom-quark mass prohibits the analogous
approximation as for the top quark, \ac{QCD} corrections to the
bottom-quark induced \ggh amplitude are known only through
\ac{NLO}, without significant progress since their original calculation
of more than 25 years ago\,\cite{Spira:1995rr}. Serious attempts to
capture the dominant logarithmic contributions of the form
$\ln\mbottom/\mhiggs$ to higher orders in perturbation theory have been
presented only recently\,\cite{Melnikov:2016emg}. Thus, also for this
source, the
\ac{LHC} Higgs Cross Section Working Group assigned an uncertainty of
roughly another 1\% to the total gluon fusion Higgs cross
section~\cite{deFlorian:2016spz}.

The total cross section at \ac{NNLO} requires the inclusion of
\three-loop virtual corrections to the \ggh amplitude (the
``Higgs-gluon form factor''), \two-loop corrections to single-real
emission, and the \one-loop double-real emission contributions which
occur for the first time at this order. The real-emission contributions
are sufficient if one aims for Higgs boson production at non-zero
transverse momenta $\pt$. In this case, top-mass effects have been
addressed by several groups
recently~\cite{Neumann:2016dny,Lindert:2018iug,Jones:2018hbb,Neumann:2018bsx}.
After estimates based on $1/\mtop$ expansions of the cross section which
indicated a break-down of this approximation for $\pt\gtrsim
150$\,GeV\,\cite{Harlander:2012hf,Neumann:2014nha}, it came as a
surprise to find the K-factor of the exact calculation to be fairly
independent of $\pt$~\cite{Lindert:2018iug,Jones:2018hbb}. This provides
yet another indication that also for the total cross section, the
\ac{QCD} corrections are well described by their heavy-top limit. Also
bottom-quark mass effects have been considered for finite
$\pt$\,\cite{Caola:2018zye,Lindert:2017pky} at this order of
perturbation theory.

Concerning the virtual corrections, it took about ten years before the
original numerical \textit{\two-loop} result for the Higgs-gluon form
factor of \citere{Spira:1995rr}, contributing to the total cross section
at \ac{NLO}, was expressed in closed analytic form using harmonic
polylogarithms\,\cite{Harlander:2005rq,Anastasiou:2006hc,Aglietti:2006tp}. The
analytic result for the \aah amplitude had been obtained
one year earlier\,\cite{Fleischer:2004vb}.

For the \three-loop form factor, only approximate results are available
up to now, most notably through expansions in the heavy-quark
limit\,\cite{Steinhauser:1996wy,Maierhofer:2012vv,Harlander:2009bw,Pak:2009bx}.
While this expansion is expected to work very well for on-shell Higgs
production mediated by a top-quark loops, it will break down for the
bottom-mediated contribution, or in cases where the Higgs is produced as
a virtual intermediate particle, for example in off-shell or
double-Higgs production. Knowledge of the general dependence of
the \three-loop Higgs-gluon form factor on the quark/Higgs mass ratio is
thus very desirable.

Very recently, the expansions in $1/\mtop$ were combined with the
leading behavior of the amplitude at the top-threshold, i.e.\ $\hat
s\approx 4\mtop^2$ (see also \citere{Grober:2017uho}) in order to
construct Pad\'e approximants for the \three-loop \ggh amplitude
which should be valid---within intrinsic Pad\'e uncertainties---for
general Higgs and quark masses~\cite{Davies:2019nhm}.

In this paper, we provide an analytic result for a subset of the virtual
\three-loop corrections, namely those involving light (massless) quark
loops in addition to the massive (top- or bottom) quark loop. Using
\ac{IBP} identities, we reduce the occurring Feynman integrals to a set of
master integrals, which we manage to solve in terms of harmonic
polylogarithms. Comparing our result to \citere{Davies:2019nhm}, we find full
agreement for this light-fermion component within the uncertainty
estimate of \citere{Davies:2019nhm}. As a byproduct of this
calculation, we also obtain the three-loop $\aah$ form factor from which
one may directly derive the exclusive photonic decay rate of the Higgs
boson through \ac{NNLO}.


\section{Calculation}
The amplitude for the processes \ggh and \aah can be parameterized with the momenta $q_{1,2}$ of the two external vector bosons as
\begin{subequations} \label{eq:Mboth}
  \begin{align}
      \begin{split}
        {\cal M}_{\ggh}^{ab;\mu\nu}
        &=
        \delta^{ab} \big[
          (q_1^\mu q_1^\nu + q_2^\mu q_2^\nu) A_{\ggh}
          + q_1^\mu q_2^\nu B_{\ggh} \\ &\qquad\qquad
          + q_2^\mu q_1^\nu C_{\ggh}
          + (q_1\cdot q_2) g^{\mu\nu} D_{\ggh}
        \big]\,,
      \end{split} \label{eq:Mggh}
      \\
      {\cal M}_{\aah}^{\mu\nu}
      &=
      (q_1^\mu q_1^\nu + q_2^\mu q_2^\nu) A_{\aah}
      + q_1^\mu q_2^\nu B_{\aah}
      + q_2^\mu q_1^\nu C_{\aah}
      + (q_1\cdot q_2) g^{\mu\nu} D_{\aah}\,, \label{eq:Maah}
  \end{align}
\end{subequations}
where we already implied Bose symmetry. Here and in what follows, $a$
and $b$ denote color indices of the adjoint representation, while $\mu$
and $\nu$ are $d$-dimensional Lorentz indices.  Because of the trivial
color structure in eq.~\eqref{eq:Mggh} which can be projected out using
$(\delta_{ab}/N_A) {\cal M}_{\ggh}^{ab;\mu\nu}$, where $N_A$ is the
number of gauge generators, we ignore the color structure in the
following and focus only on the Lorentz structure of the amplitudes.
For both amplitudes the Ward identity\footnote{In statements valid for
both amplitudes we neglect the specification \ggh or \aah in the
notation.}
\begin{equation}
  q_{1\mu} \epsilon_{j,\nu}(q_2) {\cal M}^{\mu\nu} = 0
\end{equation}
yields a constraint
\begin{equation} \label{eq:ward}
  D = -C\,.
\end{equation}
In case of photons in the external state the even stronger Ward identity $q_{1\mu} {\cal M}_{\aah}^{\mu\nu} = 0$ leads in addition to a vanishing form factor $A_{\aah}$.

For physical quantities the only contribution stems from the form factors $C$.
Therefore, the physical part of the amplitudes can be written as
\begin{subequations}
  \begin{align}
    {\cal M}_{\ggh}^{ab;\mu\nu}
    &=
    \delta^{ab} \big[
      q_2^\mu q_1^\nu
      - (q_1\cdot q_2) g^{\mu\nu}
    \big] C_{\ggh}\,,
    \\
    {\cal M}_{\aah}^{\mu\nu}
    &=
    \big[
      q_2^\mu q_1^\nu
      - (q_1\cdot q_2) g^{\mu\nu}
    \big] C_{\aah} \,.
  \end{align}
\end{subequations}
Since gluons and photons do not directly couple to the Higgs boson, the
Feynman diagrams contributing to the \ggh and \aah
amplitudes always involve at least one closed massive quark loop if
higher orders in the electroweak coupling are neglected.  In this paper,
we address the calculation of the component of the form factors $C$
which, in addition to this massive quark loop, involve a closed loop of
a light quark (assumed massless here). Since the corresponding Yukawa
coupling vanishes, the Higgs boson will still only couple to the diagram
via the massive quark loop.

In section~\ref{sect:calculation:toolchain} we describe the toolchain
used to express the contribution from light quarks to the form
factors $C$ in terms of master integrals.  In
section~\ref{sect:calculation:masters} the method for the calculation of
the master integrals is explained.

\subsection{Toolchain} \label{sect:calculation:toolchain}
\begin{figure}
  \centering
  \begin{subfigure}[b]{.18\textwidth}
    \centering
    \includegraphics[scale=.65]{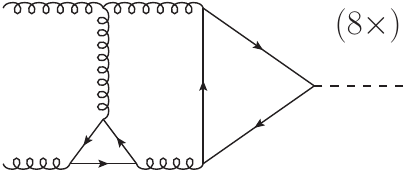}
    \caption{}
  \end{subfigure}%
  \begin{subfigure}[b]{.18\textwidth}
    \centering
    \includegraphics[scale=.65]{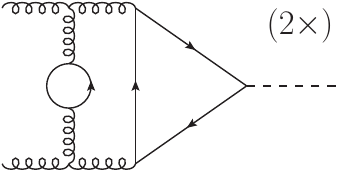}
    \caption{}
  \end{subfigure}%
  \begin{subfigure}[b]{.18\textwidth}
    \centering
    \includegraphics[scale=.65]{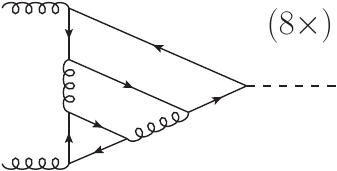}
    \caption{}
  \end{subfigure}%
  \begin{subfigure}[b]{.18\textwidth}
    \centering
    \includegraphics[scale=.65]{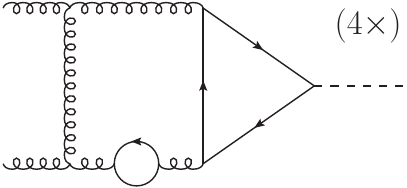}
    \caption{}
  \end{subfigure}%
  \begin{subfigure}[b]{.18\textwidth}
    \centering
    \includegraphics[scale=.65]{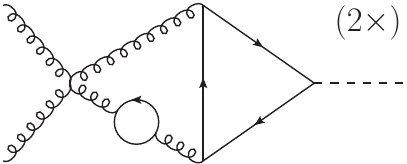}
    \caption{}
  \end{subfigure}%
  \\
  \begin{subfigure}[b]{.18\textwidth}
    \centering
    \includegraphics[scale=.65]{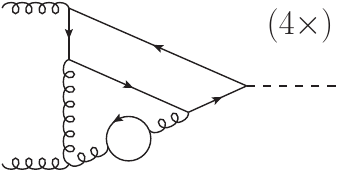}
    \caption{}
  \end{subfigure}%
  \begin{subfigure}[b]{.18\textwidth}
    \centering
    \includegraphics[scale=.65]{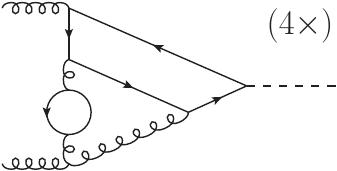}
    \caption{}
  \end{subfigure}%
  \begin{subfigure}[b]{.18\textwidth}
    \centering
    \includegraphics[scale=.65]{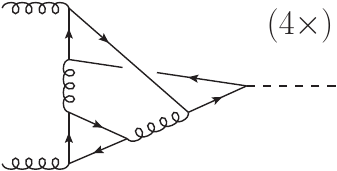}
    \caption{}
  \end{subfigure}%
  \begin{subfigure}[b]{.18\textwidth}
    \centering
    \includegraphics[scale=.65]{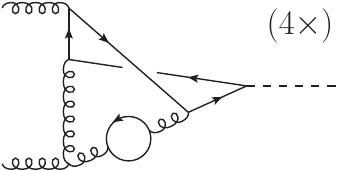}
    \caption{}
  \end{subfigure}%
  \begin{subfigure}[b]{.18\textwidth}
    \centering
    \includegraphics[scale=.65]{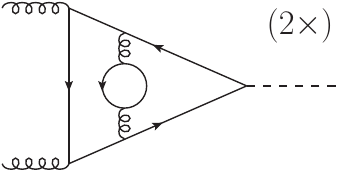}
    \caption{}
    \label{fig:feyndias:aah-first}
  \end{subfigure}%
  \\
  \begin{subfigure}[b]{.18\textwidth}
    \centering
    \includegraphics[scale=.65]{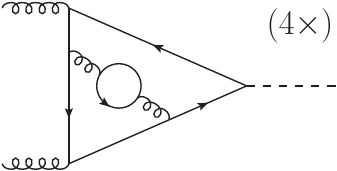}
    \caption{}
  \end{subfigure}%
  \begin{subfigure}[b]{.18\textwidth}
    \centering
    \includegraphics[scale=.65]{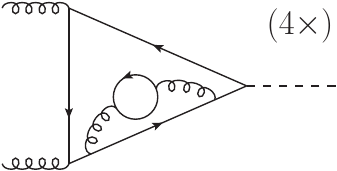}
    \caption{}
  \end{subfigure}%
  \begin{subfigure}[b]{.18\textwidth}
    \centering
    \includegraphics[scale=.65]{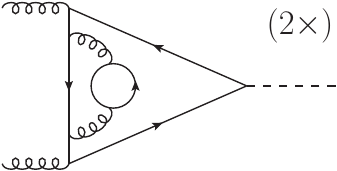}
    \caption{}
  \end{subfigure}%
  \begin{subfigure}[b]{.18\textwidth}
    \centering
    \includegraphics[scale=.65]{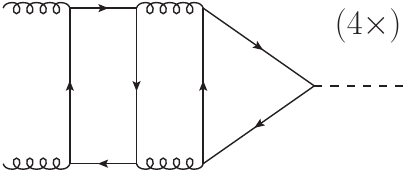}
    \caption{}
    \label{fig:feyndias:singlet1}
  \end{subfigure}%
  \begin{subfigure}[b]{.18\textwidth}
    \centering
    \includegraphics[scale=.65]{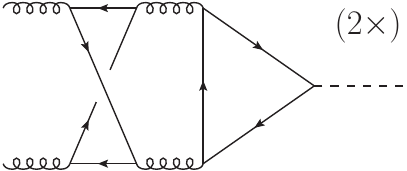}
    \caption{}
    \label{fig:feyndias:aah-last}
    \label{fig:feyndias:singlet2}
  \end{subfigure}%
\caption{Feynman diagrams for \ggh contributing to the light-quark contribution.
    Diagrams with reversed fermion flows or swapped external vector
    bosons are not shown but indicated by the multiplicity in the top
    right corner of each diagram.  Only
    diagrams \subref{fig:feyndias:aah-first}--\subref{fig:feyndias:aah-last}
    contribute to the light-quark terms of \aah with the external gluons
    replaced by photons.}
  \label{fig:feyndias}
\end{figure}%
For the calculation of the light-quark contribution to the \ggh and \aah
form factors it is required to evaluate the Feynman diagrams in
fig.~\ref{fig:feyndias}.  These Feynman diagrams are generated in a
first step using the tool \texttt{qgraf}~\cite{Nogueira:1991ex}.  After
the insertion of Feynman rules in $R_\xi$-gauge with the help of
\texttt{q2e}~\cite{Harlander:1997zb,Seidensticker:1999bb} the
diagrams are mapped to a set of seven topologies via
\texttt{exp}~\cite{Harlander:1997zb,Seidensticker:1999bb}.

A custom code for the computer algebra
system \texttt{FORM}~\cite{Vermaseren:2000nd} was written in order to
further process the output of \texttt{exp}.  We use the
projector
\begin{equation}
  {\cal P}_{\mu\nu}
  =
  \frac{
    (q_1\cdot q_2) g_{\mu\nu}
    -
    q_{2\mu} q_{1\nu}
  }{
    (q_1\cdot q_2)^2 (2-d)
  }
\end{equation}
to project out the form factor $C$ which already implies the validity of the Ward identity in eq.~\eqref{eq:ward}.
Moreover, we use
\begin{subequations}
  \begin{align}
    {\cal P}^{(A)}_{\mu\nu}
    &=
    \frac{q_{1\mu}q_{1\nu}}{(q_1\cdot q_2)^2}
    \,,\\
    {\cal P}^{(B)}_{\mu\nu}
    &=
    \frac{
      (1-d) q_{1\nu} q_{2\mu}
      -
      q_{1\mu} q_{2\nu}
      +
      (q_1\cdot q_2) g_{\mu\nu}
    }{
      (q_1\cdot q_2)^2(2-d)
    }
    \,,\\
    {\cal P}^{(C)}_{\mu\nu}
    &=
    \frac{
      -
      q_{1\nu} q_{2\mu}
      +
      (1-d) q_{1\mu} q_{2\nu}
      +
      (q_1\cdot q_2) g_{\mu\nu}
    }{
      (q_1\cdot q_2)^2(2-d)
    }
    \,,\\
    {\cal P}^{(D)}_{\mu\nu}
    &=
    \frac{
      q_{1\nu} q_{2\mu}
      +
      q_{1\mu} q_{2\nu}
      -
      (q_1\cdot q_2)
      g_{\mu\nu}
    }{
      (q_1\cdot q_2)^2(2-d)
    }
  \end{align}
\end{subequations}
to project out all the form factors in eq.~\eqref{eq:Mboth} in order to
check our calculational setup by explicitly verifying the validity of
the Ward identities, see eq.~(\ref{eq:ward}) and below.  The color
factor of each diagram is determined via the \texttt{FORM}
package \texttt{color}~\cite{vanRitbergen:1998pn}.

After projecting out the form factors, the results can be expressed in
terms of scalar Feynman integrals, which are subsequently reduced to 45
master integrals using integration-by-parts
identities~\cite{Tkachov:1981wb,Chetyrkin:1981qh} and the Laporta
algorithm~\cite{Laporta:2001dd}, implemented in the computer
program \texttt{Kira}\footnote{We note that \texttt{Kira} is also able
to completely reduce the full form factors $C$ (including the
$n_l^0$-terms) in Feynman gauge to 403 master
integrals.}~\cite{Maierhoefer:2017hyi,Maierhofer:2018gpa}.

After the reduction to master integrals the dependence on the gauge parameter $\xi$ drops out and the validity of eq.~\eqref{eq:ward} as well as $A_{\aah} = 0$ is confirmed.

\subsection{Calculation of master integrals} \label{sect:calculation:masters}
A very successful technique for the evaluation of \two-scale Feynman
integrals is based on the method of differential
equations~\cite{Kotikov:1990kg,Kotikov:1991hm,Kotikov:1991pm,Gehrmann:1999as}.
The solution of the resulting coupled system of differential equations
simplifies significantly if it can be written in the canonical form
proposed in~\citere{Henn:2013pwa}, where the right-hand side of the
system is proportional to $\epsilon=(4-d)/2$.  An algorithm to compute a
basis transformation to such a canonical form was presented by Lee
in \citere{Lee:2014ioa}. We utilize its implementation in the computer
program \texttt{epsilon}~\cite{Prausa:2017ltv} in order to evaluate the
relevant master integrals.

The class of transformations Lee's algorithm is able to find is
restricted to be rational in the kinematic variable.  Hence, a proper
choice for the kinematic variable is inevitable to obtain a canonical
form.  For our purposes, an appropriate variable is
\begin{equation}
  x = \frac{\sqrt{1-1/\tau}-1}{\sqrt{1-1/\tau}+1}\,,
\end{equation}
where $\tau = \mhiggs^2/(4\mquark^2)+i0$, with the mass $\mhiggs$ of the Higgs boson
and the mass $\mquark$ of the massive quark.

Ordering the master integrals by the number of lines in their topology
yields a block-triangular structure of the system of differential
equations.  For most applications it is sufficient to transform only the
on-diagonal blocks into the previously described canonical form.  The
differential equations for master integrals of a certain block $\vec
f(x,\epsilon)$ can then be written as
\begin{equation} \label{eq:semi-canonical}
  \frac{\partial}{\partial x} \vec f(x,\epsilon)
  =
  \epsilon M(x) \vec f(x,\epsilon)
  +
  B(x,\epsilon) \vec g(x,\epsilon)
  \,,
\end{equation}
where $\vec g(x,\epsilon)$ consists of already solved master integrals
of a lower topology, and $M(x)$ is fuchsian, i.e.\ it possesses only
simple poles in $x$.  For the master integrals entering the light-quark
contributions, these poles lie at $x=-1,0,1$. The homogeneous part
of \eqref{eq:semi-canonical} can be solved using an evolution operator
$U(x,x_0;\epsilon)$ which fulfills
\begin{equation} \label{eq:diff-U}
  \frac{\partial}{\partial x} U(x,x_0;\epsilon)
  =
  \epsilon M(x) U(x,x_0;\epsilon)
  \quad;\quad
  U(x_0,x_0;\epsilon) = \mathds1
  \,,
\end{equation}
via iterated integrations in terms of multiple polylogarithms.
This evolution operator allows expressing the full solution as
\begin{equation} \label{eq:f-int}
  \vec f(x,\epsilon)
  =
  \int_{x_0}^x \mathrm{d}x'\, U(x,x';\epsilon) B(x',\epsilon) \vec g(x',\epsilon)
  +
  U(x,x_0;\epsilon) \vec f(x_0,\epsilon)
  \,,
\end{equation}
where $\vec f(x_0,\epsilon)$ are the boundary conditions of the master integrals at $x=x_0$.
In order to simplify the integral in \eqref{eq:f-int} we chose to transform the off-diagonal blocks $B(x,\epsilon)$ of the system of differential equations into fuchsian form via \texttt{epsilon}.
Doing that ensures $\vec f(x,\epsilon)$ to be a linear combination of multiple polylogarithms without rational function prefactors in case $\vec g(x,\epsilon)$ is of this form as well.

The boundary conditions $\vec f(x_0,\epsilon)$ are calculated as an asymptotic expansion around $x_0 = 1$, which corresponds to a limit where the quark mass $\mquark$ is large compared to the Higgs mass $\mhiggs$.
For this purpose, we expand the master integrals by subgraphs~\cite{Gorishnii:1989dd,Smirnov:1990rz,Smirnov:1994tg,Smirnov:2002pj} as it is implemented in the computer program \texttt{exp}~\cite{Harlander:1997zb,Seidensticker:1999bb}.

Via the method described in this section we were able to solve not only
the 45 master integrals relevant for the light-quark contributions, but
in total 202 master integrals for the full amplitudes (including $n_l^0$
terms).  The master integrals entering the light-quark contributions
were cross checked against numerical results obtained by the
package \texttt{FIESTA}~\cite{Smirnov:2015mct}.


\section{Results}

In this section, we define the parameterization of our results, which we
have evaluated for a general gauge group with fundamental and adjoint
quadratic Casimir eigenvalues $C_F$ and $C_A$, and fundamental trace
normalization $T_F$. For \ac{QCD}, it is $C_F=4/3$, $C_A=3$, and
$T_F=1/2$. The actual analytic expressions are deferred to the appendix
for the sake of readability of the main text. In this section, we
restrict ourselves to a numerical presentation of the results.


\subsection{Results for $C_{\aah}$}
\begin{figure}
  \centering
  \begin{subfigure}[b]{.45\textwidth}
    \centering
    \includegraphics[scale=.86]{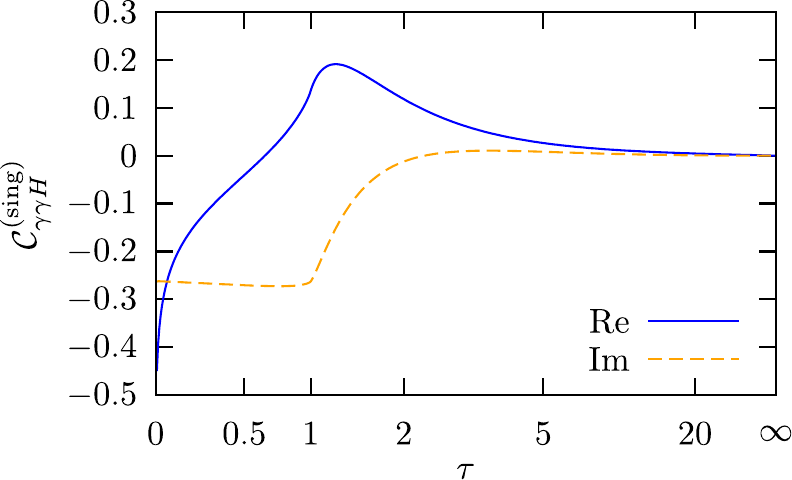}
    \caption{}
  \end{subfigure}%
  \begin{subfigure}[b]{.45\textwidth}
    \centering
    \includegraphics[scale=.86]{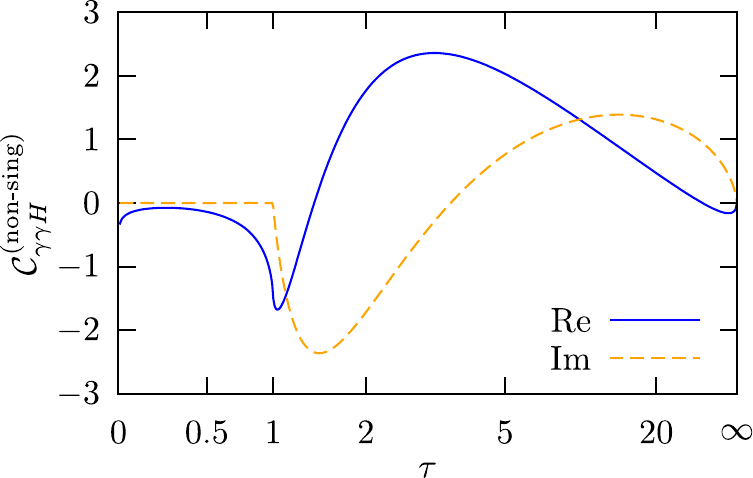}
    \caption{}
  \end{subfigure}
  \caption{\label{fig:aah}
    Singlet and non-singlet part of $C_{\aah}^{(2)}$ with the quark mass renormalized in the on-shell scheme and $\mu^2 = \mhiggs^2$.
  }
\end{figure}
The form factor $C_{\aah}$ is presented as a perturbative series in the
strong coupling constant $\alpha_s$, renormalized in
$n_l$-flavor \ac{QCD} in the \MSbar\ scheme:
\begin{equation} \label{eq:results:Caah}
  C_{\aah}
  =
  \frac1v \frac{\alpha}{\pi}
  \left[
    C_{\aah}^{(0)}
    +
    \frac{\alpha_s}\pi
    C_{\aah}^{(1)}
    +
    \left(\frac{\alpha_s}\pi\right)^2
    C_{\aah}^{(2)}
    +
    \order{\alpha_s^3}
    \right]\,,
\end{equation}
where $v$ denotes the vacuum expectation value and $\alpha$ the
electromagnetic coupling constant.  In order to fix the notation we
provide the \one-loop result as
\begin{equation}
  C_{\aah}^{(0)}
  =
  \frac{C_A Q_{\mathrm{q}}^2}{T_F} \left[
    - \frac{2x}{(1-x)^2}
    + \frac{x(1+x)^2}{(1-x)^4} H_{0,0}
    \right]
  \,,
\end{equation}
with the electric charge $Q_{\mathrm{q}}$ of the massive quark and $H_{0,0} = \ln^2(x)/2$.
The \three-loop result can be parameterized via
\begin{equation}
  C_{\aah}^{(2)}
  =
  C_{\aah}^{(2,0)}
  +
  C_A C_F Q_{\mathrm{q}}^2 n_l
  \mathcal{C}_{\aah}^{(\text{non-sing})}
  +
  C_A C_F \sum_{j=1}^{n_l} Q_j^2 \,
  \mathcal{C}_{\aah}^{(\mathrm{sing})}
  \,,
  \label{eq:results:ns}
\end{equation}
where $Q_j$ are the electric charges of the $n_l$ light quarks.  The
term $C_{\aah}^{(2,0)}$ denotes the part of the amplitude without a
massless quark loop, which we have not computed.  The contribution
stemming from a massless quark loop is split into a non-singlet part
$\mathcal{C}_{\aah}^{(\text{non-sing})}$ and a singlet part
$\mathcal{C}_{\aah}^{(\mathrm{sing})}$.  The singlet part contains the
contributions from the diagrams
figs.~\ref{fig:feyndias:singlet1}--\subref{fig:feyndias:singlet2}, where
the external photons couple to the light quark loop.

The explicit results for $\mathcal{C}_{\aah}^{(\text{non-sing})}$ and
$\mathcal{C}_{\aah}^{(\mathrm{sing})}$ are presented along with
the \two-loop results in Appendix\,\ref{sect:appdx:aah-results}, both
for an on-shell and an
\MSbar-renormalized massive quark mass. In addition, we provide them in
electronic form in an ancillary file, see Appendix~\ref{sect:appdx:anc}.
The \MSbar-renormalized result, expanded around $x=1$ up to the order
\order{(1-x)^{40}}, agrees with \citere{Maierhofer:2012vv}.

Fig.\,\ref{fig:aah} shows the real and imaginary part of the \three-loop
amplitude, separately for the singlet and the non-singlet component.


\subsection{Results for $C_{\ggh}$}
\begin{figure}
  \centering
  \begin{subfigure}[b]{.45\textwidth}
    \centering
    \includegraphics[scale=.86]{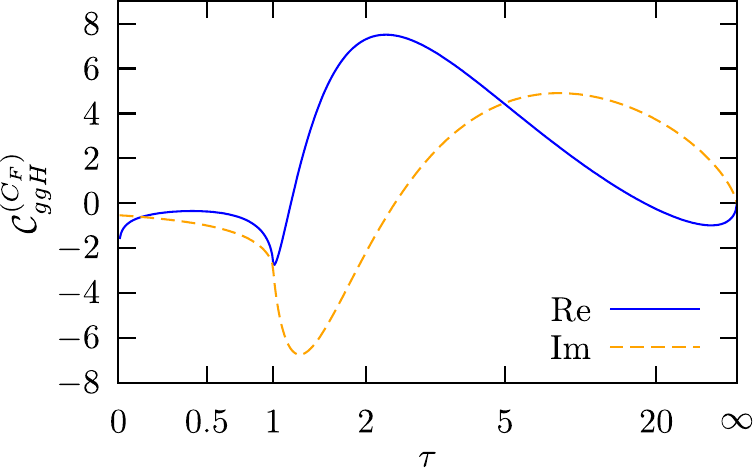}
    \caption{}
  \end{subfigure}%
  \begin{subfigure}[b]{.45\textwidth}
    \centering
    \includegraphics[scale=.86]{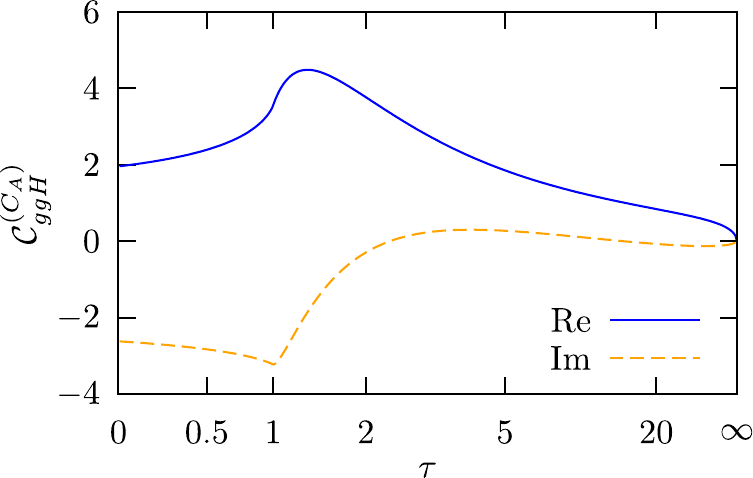}
    \caption{}
  \end{subfigure}
  \caption{\label{fig:ggh}
    Contributions to $\tilde{C}_{\ggh}^{(2)}$ separated by their color factors.
    The quark mass is renormalized in the on-shell scheme and the renormalization scale is set to $\mu^2=\mhiggs^2$.
  }
\end{figure}

The form factor $C_{\ggh}$ can also be written as a series in $\alpha_s$,
\begin{equation} \label{eq:results:Cggh}
  C_{\ggh}
  =
  \frac1v
  \frac{\alpha_s}{\pi}
  \left[
    C_{\ggh}^{(0)}
    +
    \frac{\alpha_s}\pi
    C_{\ggh}^{(1)}
    +
    \left(\frac{\alpha_s}\pi\right)^2
    C_{\ggh}^{(2)}
    +
    \order{\alpha_s^3}
    \right]\,.
\end{equation}
We provide again the \one-loop result to fix the notation,
\begin{equation}
  C_{\ggh}^{(0)} =
    T_F \left[
     -
     \frac{4x}{(1-x)^2}
     +
     \frac{2x(1 + x)^2}{(1-x)^4} H_{0,0}
    \right]
  \,.
\end{equation}
In contrast to the $C_{\aah}$ form factor, the purely virtual \ggh
result is not finite after the ultraviolet renormalization procedure.
This is due to infrared divergences which cancel against real
corrections, or are absorbed into \acp{PDF}.  In \citere{Catani:1998bh}
it is shown that the structure of these infrared divergences is
universal and can be subtracted:
\begin{subequations}
  \begin{align}
    \tilde C_{\ggh}^{(1)}
    &=
    C_{\ggh}^{(1)} - \frac12 I_g^{(1)} C_{\ggh}^{(0)}
    \,,
    \\
    \tilde C_{\ggh}^{(2)}
    &=
    C_{\ggh}^{(2)} - \frac12 I_g^{(1)} C_{\ggh}^{(1)} - \frac14 I_g^{(2)} C_{\ggh}^{(0)}
    \,.
    \label{eq:IRsubtract}
  \end{align}
\end{subequations}
The factors $I_g^{(1)}$ and $I_g^{(2)}$ are given
by~\cite{Catani:1998bh,deFlorian:2012za}
\begin{subequations}
  \begin{align}
    I_g^{(1)}
    &\equiv
    I_g^{(1)}(\epsilon)
    =
    -\left(-\frac{\mu^2}{\mhiggs^2}\right)^{\epsilon}
    \frac{e^{\epsilon\gamma_E}}{\Gamma(1-\epsilon)}
    \left[\frac{C_A}{\epsilon^2} + \frac{\beta_0}{\epsilon}\right]
    \,, \\
    \begin{split}
      I_g^{(2)}
      &=
      -
      \frac12 I_g^{(1)}(\epsilon)
      \left(
        I_g^{(1)}(\epsilon)
        +
        \frac{\beta_0}{\epsilon}
      \right)
      +
      \frac{e^{-\epsilon\gamma_E} \Gamma(1-2\epsilon)}{\Gamma(1-\epsilon)}
      \left(
        \frac{\beta_0}{\epsilon}
        +
        K
      \right)
      I_g^{(1)}(2\epsilon)
      \\ &\qquad
      +
      \left(-\frac{\mu^2}{\mhiggs^2}\right)^{2\epsilon}
      \frac{e^{\epsilon\gamma_E}}{\Gamma(1-\epsilon)}
      \frac{H_g}{2\epsilon}
      \,,
    \end{split}
  \end{align}
\end{subequations}
and
\begin{subequations}
  \begin{align}
    \beta_0
    &=
    \frac{11}6 C_A
    -
    \frac23 T_F n_l
    \,, \\
    K
    &=
    \left(\frac{67}{18} - \frac{\pi^2}6\right) C_A
    -
    \frac{10}{9} T_F n_l
    \,, \\
    H_g
    &=
    \frac{20}{27} T_F^2 n_l^2
    +
    T_F C_F n_l
    -
    \left(\frac{\pi^2}{36} + \frac{58}{27}\right) T_F n_l C_A
    +
    \left(\frac{\zeta_3}2 + \frac5{12} + \frac{11\pi^2}{144}\right) C_A^2
    \,.
  \end{align}
\end{subequations}
\begin{figure}
  \centering
  \begin{subfigure}[b]{.45\textwidth}
    \centering
    \includegraphics[scale=.86]{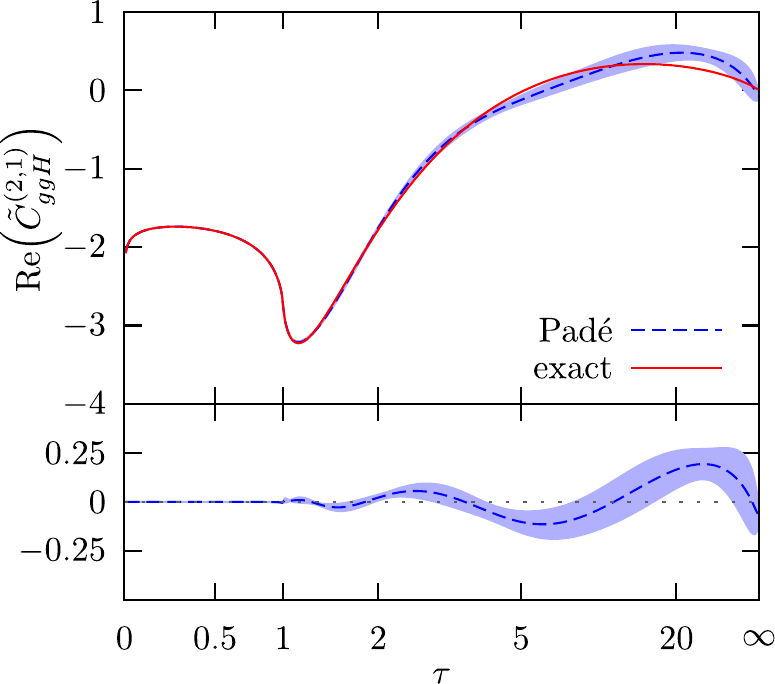}
    \caption{}
  \end{subfigure}%
  \begin{subfigure}[b]{.45\textwidth}
    \centering
    \includegraphics[scale=.86]{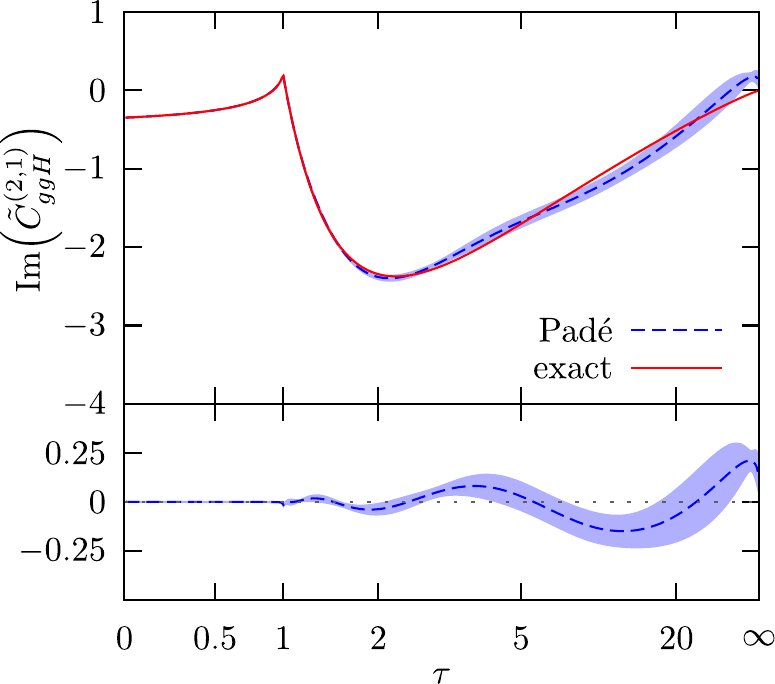}
    \caption{}
  \end{subfigure}
  \caption{  \label{fig:compare}
    Real and imaginary part of $\tilde C^{(2,1)}_{\ggh}$ with the quark
  mass renormalized on-shell, $\mu^2=-\mhiggs^2$, and the color factors
  set to their $SU(3)$
  values.
    The red (solid) line is the result given in eqs.~(\ref{eq:Cggh21},\ref{eq:Cggh21cF},\ref{eq:Cggh21cA}).
    The blue (dashed) line shows the Pad\'e approximation found
  in \citere{Davies:2019nhm}; the associated uncertainty estimate is
  indicated by the blue shaded band.
    The lower panel shows the difference between the Pad\'e approximation and our result.
  }
\end{figure}
The finite \three-loop terms can be parameterized as
\begin{equation}
  \tilde C_{\ggh}^{(2)}
  =
  \tilde C_{\ggh}^{(2,0)}
  +
  n_l
  \tilde C_{\ggh}^{(2,1)}
  +
  n_l^2
  \tilde C_{\ggh}^{(2,2)}
  \,,
  \label{eq:cgghnlsplit}
\end{equation}
where $\tilde C_{\ggh}^{(2,0)}$ denotes the contribution of Feynman
diagrams without a light quark loop.  Since at the \three-loop level
there are no diagrams with more than two closed quark loops, the $\tilde
C_{\ggh}^{(2,2)}$ contribution originates solely from the subtraction
terms of eq.~(\ref{eq:IRsubtract}).  We further decompose $\tilde
C_{\ggh}^{(2,1)}$ into its contributions to different color factors,
\begin{equation} \label{eq:Cggh21}
  \tilde C_{\ggh}^{(2,1)}
  =
  T_F^2 C_F \mathcal{C}_{\ggh}^{(C_F)}
  +
  T_F^2 C_A \mathcal{C}_{\ggh}^{(C_A)}
  \,.
\end{equation}
Explicit results for $\mathcal{C}_{\ggh}^{(C_F)}$ and
$\mathcal{C}_{\ggh}^{(C_A)}$, where the quark mass again has been
renormalized both in the \MSbar and the on-shell scheme, are given along
with the \two-loop results in Appendix~\ref{sect:appdx:ggh-results} and
in an ancillary file, see Appendix~\ref{sect:appdx:anc}.  The result
renormalized in the on-shell scheme, expanded around $x=1$ up
to \order{(1-x)^{12}}, agrees
with Refs.~\cite{Davies:2019nhm,Harlander:2009mq,Pak:2009bx}. Also all
terms of the threshold expansion given in Refs.~\cite{Grober:2017uho,Davies:2019nhm} could be reproduced.
The real and imaginary part of the result is shown in fig.\,\ref{fig:ggh}.

Fig.\,\ref{fig:compare} compares these results to the semi-numerical
approximation of \citere{Davies:2019nhm}.\footnote{ In accordance
with \citere{Davies:2019nhm}, we set $\mu^2=-\mhiggs^2$ in
fig.\,\ref{fig:compare}.} The latter is associated with a systematic
uncertainty due to the approximation procedure, which is dominated at
large $\tau$ by the absence of any input from this kinematical region
into the Pad\'e approximants. Our results indeed confirm the associated
uncertainty estimate for the light-quark terms up to rather large values
of~$\tau$ (corresponding to large Higgs masses/virtualities or small
quark masses).



\section{Conclusions}

The Higgs-gluon form factor is an essential component for the
theoretical description of Higgs physics at hadron colliders. It enters
the total cross section for single and double Higgs production at the
\ac{LHC}, for example. In \ac{QCD}, it involves a massive quark loop
which mediates the Higgs-gluon/photon coupling.  Until recently, the
\three-loop \ggh form factor has been known only in the limit of a very
heavy mediating quark. This has restricted its applicability to top-quark
mediated on-shell (or not-too off-shell) single production of the
\ac{SM}-like Higgs boson. Even there, the lack of an exact result
implied non-negligible uncertainties. In cases such as double-Higgs
production, off-shell or bottom-quark mediated single-Higgs production,
or the production of heavy \ac{BSM} Higgs bosons, the expansion fails
and one had to resort to the \ac{NLO} result.

In this paper, we provided an analytic \three-loop result for the
component of the gluon-Higgs form factor which, in addition to the
massive quark loop, involves a closed massless quark loop. We showed
that it can be computed in closed form for a general quark mass, and
presented it in terms of harmonic polylogarithms. Comparison of this
result to a recent semi-numerical evaluation of the full Higgs-gluon
form factor shows very good agreement at the level of the estimated
numerical uncertainties. As a byproduct of our calculation, we also
presented the analogous component of the amplitude for the decay rate
of the Higgs boson into photons.

A large portion of our technical setup is applicable also to the full
form factor. However, the calculations are much more expensive than for
the light-quark terms considered here. Moreover, one encounters elliptic
integrals which cannot be expressed in terms of harmonic
polylogarithms. A fully analytical result for the \three-loop form factor
therefore requires further efforts.


\paragraph{Acknowledgments.}
We would like to thank Florian Herren and Philipp Maierh\"ofer for
helpful discussions, and Matthias Steinhauser for useful comments on the
manuscript. This work was supported by Deutsche Forschungsgemeinschaft
(DFG) through the Collaborative Research Center TRR 257 ``Particle
Physics Phenomenology after the Higgs Discovery''.
J.U. received funding from the European Research Council (ERC) under the
European Unions Horizon 2020 research and innovation programme under grant
agreement no. 647356 (CutLoops).
The authors acknowledge support by the state of Baden-Württemberg through bwHPC
and the German Research Foundation (DFG) through grant no INST 39/963-1 FUGG.
Parts of the computing resources were granted by RWTH Aachen University under
project rwth0119.
We would also like to thank Peter Uwer and his group ``Phenomenology of Elementary
Particle Physics beyond the Standard Model'' at Humboldt-Universit\"at zu Berlin
for providing computer resources. The Feynman diagrams in this article have been
drawn with \texttt{JaxoDraw}~\cite{Binosi:2008ig} based on \texttt{Axodraw}~\cite{Vermaseren:1994je}.


\appendix


\section{Results for $C_{\aah}$} \label{sect:appdx:aah-results}

In this appendix, we provide explicit formulas for the \two-loop result
and the newly computed light-quark contributions to the \three-loop
result of the $\aah$ form factor, cf.\
eqs.~(\ref{eq:results:Caah},\ref{eq:results:ns}). We use
$H_{\vec{a}} \equiv H(\vec a;x)$ to denote harmonic
polylogarithms~\cite{Remiddi:1999ew,Maitre:2005uu},
$\zeta_n \equiv \sum_{j=1}^\infty j^{-n}$ for Riemann's zeta function,
and the short-hand notation $L_\mu \equiv \ln(\mu^2/\mquark^2)$ with the
renormalization scale $\mu$.

The \two-loop result has been know for about 15
years\,\cite{Fleischer:2004vb}. In our notation, it reads
{\allowdisplaybreaks\begin{align}
  C_{\aah}^{(1)}
  &=
  \frac{C_A C_F Q_{\mathrm{q}}^2}{T_F} \Big\{
-\frac{5 x}{(1-x)^2}
+\frac{x \left(1-14 x+x^2\right)}{(1-x)^4}
\zeta_{3}-\frac{3 x (1+x)}{(1-x)^3}
H_{0}+\frac{6 x^2}{(1-x)^4}
H_{0,0} \nonumber \\ &\hspace{6em}
-\frac{x \left(5-6 x+5 x^2\right)}{(1-x)^4}
H_{1,0,0}+\frac{x \left(3+25 x-7 x^2+3 x^3\right)}{2 (1-x)^5}
H_{0,0,0} \nonumber \\ &\hspace{6em}
+\frac{x (1+x)^2}{6 (1-x)^4}\Big[
-\pi ^2 H_{0}-24 H_{0,-1,0}+6 H_{0,1,0}\Big]+\frac{x (1+x) \left(1+x^2\right)}{60 (1-x)^5}
 \nonumber \\ &\hspace{6em}
\times\Big[
-3 \pi ^4-20 \pi ^2 H_{0,0}+240 H_{0,-1,0,0}-480 H_{0,0,-1,0}-30 H_{0,0,0,0} \nonumber \\ &\hspace{6em}\quad
+120 H_{0,0,1,0}-420 H_{0,1,0,0}-240 H_{0} \zeta_{3}\Big]

  \Big\}
  \,, \\
  \overline{C}_{\aah}^{(1)}
  &=
  C_{\aah}^{(1)}
  +
  \Delta C_{\aah}^{(1)}
  \,, \\
  \Delta C_{\aah}^{(1)}
  &=
  \frac{C_A C_F Q_{\mathrm{q}}^2}{T_F} \Big\{
\frac{x}{(1-x)^2}\Big[
-4-3 L_\mu\Big]+\frac{x (1+x)}{2 (1-x)^3}\Big[
4 H_{0}+3 H_{0} L_\mu\Big] \nonumber \\ &\hspace{6em}
+\frac{x \left(1+6 x+x^2\right)}{2 (1-x)^4}\Big[
4 H_{0,0}+3 H_{0,0} L_\mu\Big]

  \Big\}
  \,,
\end{align}}%
where $C_{\aah}^{(1)}$\,($\overline{C}_{\aah}^{(1)}$) is the result for a quark mass renormalized in the on-shell\,(\MSbar) scheme.

At three loops, the non-singlet contribution, with the quark mass
renormalized in the on-shell scheme, is given by
{\allowdisplaybreaks\begin{align}
\mathcal{C}_{\aah}^{(\text{non-sing})} &=
-\frac{(5-7 x) x}{18 (1-x)^3}
\pi ^2-\frac{x \left(1-14 x+x^2\right)}{3 (1-x)^4}
L_\mu \zeta_{3}+\frac{x \left(13+118 x+13 x^2\right)}{9 (1-x)^4}
\zeta_{3} \nonumber \\ &\quad
+\frac{x \left(63-85 x+73 x^2+21 x^3\right)}{1080 (1-x)^5}
\pi ^4-\frac{4 x \left(1-4 x+x^2\right)}{3 (1-x)^4}
H_{1} \zeta_{3}+\frac{x \left(17+10 x-x^2\right)}{54 (1-x)^4}
 \nonumber \\ &\quad
\times \pi ^2 H_{0}+\frac{2 x \left(16+19 x-20 x^2+13 x^3\right)}{9 (1-x)^5}
H_{0} \zeta_{3}-\frac{2 x^2}{(1-x)^4}
H_{0,0} L_\mu \nonumber \\ &\quad
+\frac{x \left(3-14 x-10 x^2\right)}{3 (1-x)^4}
H_{0,0}+\frac{x \left(47+95 x-115 x^2+5 x^3\right)}{108 (1-x)^5}
\pi ^2 H_{0,0} \nonumber \\ &\quad
-\frac{2 x \left(7+29 x+7 x^2\right)}{9 (1-x)^4}
H_{0,1,0}-\frac{x \left(33+499 x-265 x^2-99 x^3\right)}{36 (1-x)^5}
H_{0,0,0} \nonumber \\ &\quad
-\frac{x \left(3+25 x-7 x^2+3 x^3\right)}{6 (1-x)^5}
H_{0,0,0} L_\mu+\frac{4 x \left(11+34 x+11 x^2\right)}{9 (1-x)^4}
H_{0,-1,0} \nonumber \\ &\quad
-\frac{2 x \left(3-2 x+3 x^2\right)}{3 (1-x)^4}
H_{1,1,0,0}-\frac{x \left(2+62 x+17 x^2+17 x^3\right)}{9 (1-x)^5}
H_{0,0,0,0} \nonumber \\ &\quad
-\frac{x \left(47+119 x-67 x^2+29 x^3\right)}{18 (1-x)^5}
H_{0,0,1,0}+\frac{8 x (1+x) \left(2+6 x-7 x^2\right)}{9 (1-x)^5}
H_{0,-1,0,0} \nonumber \\ &\quad
+\frac{x \left(17-42 x+17 x^2\right)}{6 (1-x)^4}
H_{1,0,0,0}+\frac{4 x \left(19+31 x-23 x^2+13 x^3\right)}{9 (1-x)^5}
H_{0,0,-1,0} \nonumber \\ &\quad
+\frac{2 x \left(10-11 x-11 x^2+34 x^3\right)}{9 (1-x)^5}
H_{0,1,0,0}+\frac{x (1+x)}{6 (1-x)^3}\Big[
23 H_{0}-16 H_{-1,0}+10 H_{1,0} \nonumber \\ &\quad\quad
+6 H_{0} L_\mu\Big]+\frac{x \left(5-6 x+5 x^2\right)}{9 (1-x)^4}\Big[
5 H_{1,0,0}+3 H_{1,0,0} L_\mu\Big]+\frac{x}{18 (1-x)^2}\Big[
101 \nonumber \\ &\quad\quad
-\pi ^2 H_{1,0}-96 H_{1,0,-1,0}+42 H_{1,0,1,0}+30 L_\mu\Big]+\frac{x (1+x)^2}{18 (1-x)^4}\Big[
2 \pi ^2 H_{0,1} \nonumber \\ &\quad\quad
-96 H_{0,-1,-1,0}+48 H_{0,-1,1,0}+48 H_{0,1,-1,0}-12 H_{0,1,1,0}+\pi ^2 H_{0} L_\mu-6 H_{0,1,0} L_\mu \nonumber \\ &\quad\quad
+24 H_{0,-1,0} L_\mu\Big]+\frac{x (1+x) \left(1+x^2\right)}{540 (1-x)^5}\Big[
-18 \pi ^4 H_{-1}+14 \pi ^4 H_{0}-120 \pi ^2 H_{-1,0,0} \nonumber \\ &\quad\quad
+120 \pi ^2 H_{0,0,0}+120 \pi ^2 H_{0,0,1}+1440 H_{-1,0,-1,0,0}-2880 H_{-1,0,0,-1,0}+9 \pi ^4 L_\mu \nonumber \\ &\quad\quad
-180 H_{-1,0,0,0,0}+720 H_{-1,0,0,1,0}-2520 H_{-1,0,1,0,0}+1440 H_{0,-1,-1,0,0}-30 \pi ^2 \zeta_{3} \nonumber \\ &\quad\quad
+1440 H_{0,-1,0,-1,0}-1800 H_{0,-1,0,0,0}-720 H_{0,-1,0,1,0}-5760 H_{0,0,-1,-1,0} \nonumber \\ &\quad\quad
+2880 H_{0,0,-1,0,0}+2880 H_{0,0,-1,1,0}+2880 H_{0,0,0,-1,0}+270 H_{0,0,0,0,0}+2520 \zeta_{5} \nonumber \\ &\quad\quad
-720 H_{0,0,0,1,0}+2880 H_{0,0,1,-1,0}-720 H_{0,0,1,0,0}-720 H_{0,0,1,1,0}+3060 H_{0,1,0,0,0} \nonumber \\ &\quad\quad
-4320 H_{0,1,0,-1,0}+2160 H_{0,1,0,1,0}-360 H_{0,1,1,0,0}+60 \pi ^2 H_{0,0} L_\mu+90 H_{0,0,0,0} L_\mu \nonumber \\ &\quad\quad
-720 H_{0,-1,0,0} L_\mu+1440 H_{0,0,-1,0} L_\mu-360 H_{0,0,1,0} L_\mu+1260 H_{0,1,0,0} L_\mu \nonumber \\ &\quad\quad
-1440 H_{-1,0} \zeta_{3}-360 H_{0,-1} \zeta_{3}+1080 H_{0,0} \zeta_{3}-1800 H_{0,1} \zeta_{3}+720 H_{0} L_\mu \zeta_{3}\Big]

  \,.
\end{align}}%
The singlet contribution is
{\allowdisplaybreaks\begin{align}
\mathcal{C}_{\aah}^{(\text{sing})} &=
-\frac{x \left(37-243 x+249 x^2-15 x^3\right)}{6 (1-x)^5}
\zeta_{3}+\frac{5 x \left(1-22 x+x^2\right)}{3 (1-x)^4}
\zeta_{5} \nonumber \\ &\quad
+\frac{x \left(3+44 x-68 x^2+20 x^3+15 x^4\right)}{1080 (1-x)^6}
\pi ^4-\frac{39 x^2}{2 (1-x)^3}
H_{0}-\frac{x^2 \left(4-x+11 x^2\right)}{18 (1-x)^5}
\pi ^2 H_{0} \nonumber \\ &\quad
+\frac{3 x}{2 (1-x)^2}\Big[
2+13 H_{1}\Big]+\frac{4 x \left(1+3 x+x^2\right)}{(1-x)^4}
H_{1} \zeta_{3}+\frac{2 x^2 \left(14-20 x-4 x^2+3 x^3\right)}{3 (1-x)^6}
 \nonumber \\ &\quad
\times H_{0} \zeta_{3}-\frac{4 x \left(1+8 x+x^2\right)}{3 (1-x)^4}
H_{1,0} \zeta_{3}+\frac{x^2 (35+66 x)}{6 (1-x)^4}
H_{0,0}+\frac{x^2 \left(2-5 x+2 x^2-6 x^3\right)}{9 (1-x)^6}
 \nonumber \\ &\quad
\times \pi ^2 H_{0,0}-\frac{x^2 \left(22-145 x-7 x^2\right)}{6 (1-x)^5}
H_{0,0,0}-\frac{2 x \left(1-x+x^2\right)}{9 (1-x)^4}
\pi ^2 H_{1,0,0} \nonumber \\ &\quad
+\frac{x \left(13-266 x+13 x^2\right)}{6 (1-x)^4}
H_{1,0,0}+\frac{2 x \left(4-22 x+25 x^2+7 x^3\right)}{3 (1-x)^5}
H_{0,-1,0} \nonumber \\ &\quad
+\frac{x \left(1-3 x+9 x^2+21 x^3\right)}{3 (1-x)^5}
H_{0,0,1}-\frac{2 x \left(1+2 x-7 x^2\right)}{(1-x)^5}
H_{0,1,0,0}-\frac{2 x \left(1+10 x+x^2\right)}{(1-x)^4}
 \nonumber \\ &\quad
\times H_{1,1,0,0}+\frac{2 x^2 (1+3 x)}{(1-x)^5}
H_{0,0,1,0}+\frac{2 x \left(1-6 x-2 x^2\right)}{(1-x)^4}
H_{1,0,0,0} \nonumber \\ &\quad
+\frac{x^2 \left(2+7 x-10 x^2-6 x^3\right)}{3 (1-x)^6}
H_{0,0,0,0}+\frac{x \left(1-10 x+x^2\right)}{(1-x)^4}
H_{0,1,0,0,0}+\frac{x (1+x)}{36 (1-x)^3}
 \nonumber \\ &\quad
\times\Big[
-7 \pi ^2+444 H_{-1,0}+156 H_{0,1}-108 H_{1,0}\Big]+\frac{x (1+x) \left(11-8 x+11 x^2\right)}{18 (1-x)^5}\Big[
\pi ^2 H_{-1} \nonumber \\ &\quad\quad
-12 H_{-1,-1,0}-12 H_{-1,0,1}\Big]+\frac{x \left(1+x^2\right)}{3 (1-x)^4}\Big[
12 H_{1,0,-1,0}+12 H_{1,0,0,1}+\pi ^2 \zeta_{3}\Big] \nonumber \\ &\quad
+\frac{4 x^2 \left(1+11 x-11 x^2+6 x^3\right)}{3 (1-x)^6}\Big[
H_{0,0,-1,0}+H_{0,0,0,1}\Big] \nonumber \\ &\quad
+\frac{x \left(3-4 x+16 x^2-4 x^3+3 x^4\right)}{9 (1-x)^6}\Big[
\pi ^2 H_{0,-1}-12 H_{0,-1,-1,0}-12 H_{0,-1,0,1}\Big] \nonumber \\ &\quad
+\frac{x \left(1-4 x+x^2\right)}{270 (1-x)^4}\Big[
\pi ^4 H_{1}-270 H_{0,1,0}+720 H_{1,0,0,-1,0}+720 H_{1,0,0,0,1}\Big]+\frac{x (1+x)^2}{9 (1-x)^4}
 \nonumber \\ &\quad
\times \Big[
-3 \pi ^2 H_{1,0}+2 \pi ^2 H_{0,0,-1}+4 \pi ^2 H_{1,0,-1}-24 H_{0,0,-1,-1,0}-24 H_{0,0,-1,0,1} \nonumber \\ &\quad\quad
-3 H_{0,0,1,0,0}-48 H_{1,0,-1,-1,0}-48 H_{1,0,-1,0,1}-6 H_{1,0,0,0,0}\Big]+\frac{2 x^2}{3 (1-x)^4}\Big[
\pi ^2 H_{0,1,0} \nonumber \\ &\quad\quad
+2 \pi ^2 H_{1,1,0} -6 H_{1,0,1,0}-24 H_{0,1,0,-1,0}-24 H_{0,1,0,0,1}-6 H_{0,1,0,1,0}-12 H_{0,1,1,0,0} \nonumber \\ &\quad\quad
-6 H_{1,0,0,1,0}-12 H_{1,0,1,0,0}-48 H_{1,1,0,-1,0}-18 H_{1,1,0,0,0}-48 H_{1,1,0,0,1} \nonumber \\ &\quad\quad
-12 H_{1,1,0,1,0}-12 H_{0,1} \zeta_{3}-24 H_{1,1,1,0,0}-24 H_{1,1} \zeta_{3}\Big]

  \,.
\end{align}}%
The results with an \MSbar renormalized quark mass can be written as
\begin{align}
  \overline{\mathcal{C}}_{\aah}^{(\text{non-sing})}
  &=
  \mathcal{C}_{\aah}^{(\text{non-sing})}
  +
  \Delta \mathcal{C}_{\aah}^{(\text{non-sing})}
  \\
  \overline{\mathcal{C}}_{\aah}^{(\text{sing})}
  &=
  \mathcal{C}_{\aah}^{(\text{sing})}
  \,,
\end{align}
where
\begin{align}
\Delta \mathcal{C}_{\aah}^{(\text{non-sing})} &= 
\frac{x}{24 (1-x)^2}\Big[
71+8 \pi ^2+52 L_\mu+12 L_\mu^{2}\Big]+\frac{x (1+x)}{48 (1-x)^3}\Big[
-71 H_{0}-8 \pi ^2 H_{0} \nonumber \\ &\quad\quad
-52 H_{0} L_\mu-12 H_{0} L_\mu^{2}\Big]+\frac{x \left(1+6 x+x^2\right)}{48 (1-x)^4}\Big[
-71 H_{0,0}-8 \pi ^2 H_{0,0}-52 H_{0,0} L_\mu \nonumber \\ &\quad\quad
-12 H_{0,0} L_\mu^{2}\Big]

  \,.
\end{align}
Note that the singlet component appears for the first time at
the \three-loop order and is therefore renormalization scheme
independent.


\section{Results for $C_{\ggh}$} \label{sect:appdx:ggh-results}

In this appendix, we provide explicit formulas for the \two-loop and the
newly computed light-quark contribution to the \three-loop result of the
$\ggh$ form factor, cf.\
eqs.~(\ref{eq:results:Cggh},\ref{eq:Cggh21}). The notation is the same
as in Appendix~\ref{sect:appdx:aah-results}.

Again, the \two-loop result has been known for about 15
years~\cite{Harlander:2005rq,Aglietti:2006tp,Anastasiou:2006hc}. In our
notation, it reads
{\allowdisplaybreaks\begin{align}
  \tilde C_{\ggh}^{(1)}
  &=
  T_F C_A
  \Big\{
\frac{2 x (3+x) (1+3 x)}{(1-x)^4}
\zeta_{3}+\frac{x}{3 (1-x)^2}\Big[
-18-11 H_{0}-22 H_{1}-11 L_\mu\Big] \nonumber \\ &\hspace{4.2em}
-\frac{x \left(7+38 x+7 x^2\right)}{3 (1-x)^4}
H_{1,0,0}+\frac{x \left(11+15 x+21 x^2+x^3\right)}{2 (1-x)^5}
H_{0,0,0}+\frac{x (1+x)^2}{90 (1-x)^4}
 \nonumber \\ &\hspace{4.2em}\quad
\times\Big[8 \pi ^4+90 H_{0,0}+30 \pi ^2 H_{0,0}+60 \pi ^2 H_{1,0}+330 H_{0,0,1}+330 H_{0,1,0} \nonumber \\ &\hspace{4.2em}\quad
+90 H_{0,0,0,0}+360 H_{0,0,-1,0}+720 H_{1,0,-1,0}-360 H_{1,0,0,0}+165 H_{0,0} L_\mu \nonumber \\ &\hspace{4.2em}\quad
+540 H_{0} \zeta_{3}+1080 H_{1} \zeta_{3}\Big]

  \Big\}
  \nonumber \\ &\quad
  +
  T_F C_F
  \Big\{
-\frac{10 x}{(1-x)^2}
+\frac{2 x \left(1-14 x+x^2\right)}{(1-x)^4}
\zeta_{3}-\frac{6 x (1+x)}{(1-x)^3}
H_{0}+\frac{12 x^2}{(1-x)^4}
H_{0,0} \nonumber \\ &\hspace{5.4em}
-\frac{2 x \left(5-6 x+5 x^2\right)}{(1-x)^4}
H_{1,0,0}+\frac{x \left(3+25 x-7 x^2+3 x^3\right)}{(1-x)^5}
H_{0,0,0} \nonumber \\ &\hspace{5.4em}
+\frac{x (1+x)^2}{3 (1-x)^4}\Big[
-\pi ^2 H_{0}-24 H_{0,-1,0}+6 H_{0,1,0}\Big]+\frac{x (1+x) \left(1+x^2\right)}{30 (1-x)^5}\Big[
-3 \pi ^4 \nonumber \\ &\hspace{5.4em}\quad
-20 \pi ^2 H_{0,0}+240 H_{0,-1,0,0}-480 H_{0,0,-1,0}-30 H_{0,0,0,0}+120 H_{0,0,1,0} \nonumber \\ &\hspace{5.4em}\quad
-420 H_{0,1,0,0}-240 H_{0} \zeta_{3}\Big]

  \Big\}
  \nonumber \\ &\quad
  + n_l T_F^2
  \Big\{
\frac{4 x}{3 (1-x)^2}\Big[
H_{0}+2 H_{1}+L_\mu\Big]+\frac{2 x (1+x)^2}{3 (1-x)^4}\Big[
-3 H_{0,0,0}-2 H_{0,0,1}-2 H_{0,1,0} \nonumber \\ &\hspace{5.5em}\quad
-2 H_{1,0,0}-H_{0,0} L_\mu\Big]

  \Big\}
  \,, \\
  \tilde{\overline{C}}_{\ggh}^{(1)}
  &=
  \tilde C_{\ggh}^{(1)}
  +
  \Delta \tilde C_{\ggh}^{(1)}
  \,, \\
  \Delta \tilde C_{\ggh}^{(1)}
  &=
  T_F C_F \Big\{
\frac{2 x}{(1-x)^2}\Big[
-4-3 L_\mu\Big]+\frac{x (1+x)}{(1-x)^3}\Big[
4 H_{0}+3 H_{0} L_\mu\Big]
 \nonumber \\ &\hspace{4em}\quad
+\frac{x \left(1+6 x+x^2\right)}{(1-x)^4}\Big[4 H_{0,0}+3 H_{0,0} L_\mu\Big]

  \Big\}
  \,,
\end{align}}%
where again the symbols with (without) a bar on top denote the results
for an \MSbar\,(on-shell) renormalized quark mass.

The contribution to the \three-loop \ggh form factor proportional to the
color factor $C_F$, cf.\ eq.~(\ref{eq:Cggh21}), is given by
{\allowdisplaybreaks\begin{align} 
\mathcal{C}_{\ggh}^{(C_F)} &= 
-\frac{(17-7 x) x}{18 (1-x)^3}
\pi ^2-\frac{4 x \left(1-14 x+x^2\right)}{3 (1-x)^4}
L_\mu \zeta_{3}-\frac{x \left(85-939 x+957 x^2-19 x^3\right)}{9 (1-x)^5}
\zeta_{3} \nonumber \\ &\quad
+\frac{2 (1-3 x) x \left(19-44 x-3 x^2\right)}{3 (1-x)^5}
\zeta_{5}+\frac{(5-7 x) x \left(1+x^2\right)}{9 (1-x)^5}
\pi ^2 \zeta_{3} \nonumber \\ &\quad
+\frac{x \left(33-52 x+45 x^2-16 x^3-3 x^4\right)}{270 (1-x)^6}
\pi ^4+\frac{x}{9 (1-x)^2}\Big[
155+429 H_{1}+69 L_\mu\Big] \nonumber \\ &\quad
+\frac{(36-107 x) x}{3 (1-x)^3}
H_{0}+\frac{4 x \left(3+40 x+3 x^2\right)}{3 (1-x)^4}
H_{1} \zeta_{3}+\frac{x \left(17-19 x-8 x^2-32 x^3\right)}{27 (1-x)^5}
\pi ^2 H_{0} \nonumber \\ &\quad
+\frac{2 x \left(5+2 x+7 x^2+4 x^3\right)}{135 (1-x)^5}
\pi ^4 H_{1}+\frac{2 x \left(29+138 x-288 x^2+90 x^3-11 x^4\right)}{9 (1-x)^6}
H_{0} \zeta_{3} \nonumber \\ &\quad
-\frac{x \left(1+18 x+x^2\right)}{2 (1-x)^4}
H_{0,0} L_\mu-\frac{x \left(5+6 x+5 x^2\right)}{9 (1-x)^4}
\pi ^2 H_{1,0}-\frac{4 x \left(1+13 x-11 x^2+x^3\right)}{3 (1-x)^5}
 \nonumber \\ &\quad\quad
H_{0,1} \zeta_{3}+\frac{x \left(18+7 x+34 x^2\right)}{3 (1-x)^4}
H_{0,0}+\frac{8 x \left(1-5 x+9 x^2+3 x^3\right)}{3 (1-x)^5}
H_{1,0} \zeta_{3} \nonumber \\ &\quad
+\frac{x \left(59+72 x-294 x^2+144 x^3-65 x^4\right)}{54 (1-x)^6}
\pi ^2 H_{0,0}+\frac{2 x \left(3-4 x+16 x^2-4 x^3+3 x^4\right)}{9 (1-x)^6}
 \nonumber \\ &\quad\quad
\pi ^2 H_{0,-1}-\frac{x \left(55+134 x+55 x^2\right)}{9 (1-x)^4}
H_{0,1,0}-\frac{x \left(30+437 x-689 x^2-84 x^3\right)}{9 (1-x)^5}
H_{0,0,0} \nonumber \\ &\quad
-\frac{x \left(1+33 x-45 x^2-45 x^3\right)}{3 (1-x)^5}
H_{0,0,1}-\frac{2 x \left(3+25 x-7 x^2+3 x^3\right)}{3 (1-x)^5}
H_{0,0,0} L_\mu \nonumber \\ &\quad
+\frac{4 x^2 \left(3-x+2 x^2\right)}{9 (1-x)^5}
\pi ^2 H_{1,0,0}+\frac{4 x \left(5-6 x+5 x^2\right)}{3 (1-x)^4}
H_{1,0,0} L_\mu \nonumber \\ &\quad
+\frac{4 x \left(20-237 x+20 x^2\right)}{9 (1-x)^4}
H_{1,0,0}+\frac{4 x \left(34-20 x+29 x^2-x^3\right)}{9 (1-x)^5}
H_{0,-1,0} \nonumber \\ &\quad
-\frac{x \left(77+245 x-229 x^2+35 x^3\right)}{9 (1-x)^5}
H_{0,0,1,0}-\frac{2 x \left(3+18 x-76 x^2+54 x^3-27 x^4\right)}{3 (1-x)^6}
 \nonumber \\ &\quad\quad
H_{0,0,0,1}-\frac{2 x \left(20+186 x-258 x^2+90 x^3-17 x^4\right)}{9 (1-x)^6}
H_{0,0,0,0} \nonumber \\ &\quad
-\frac{8 x \left(1-4 x+20 x^2-4 x^3+x^4\right)}{3 (1-x)^6}
H_{0,-1,0,1}-\frac{8 x \left(7-4 x+8 x^2-4 x^3+7 x^4\right)}{3 (1-x)^6}
 \nonumber \\ &\quad\quad
H_{0,-1,-1,0}+\frac{16 x (1+x) \left(5+6 x-10 x^2\right)}{9 (1-x)^5}
H_{0,-1,0,0}+\frac{16 x \left(1-10 x+x^2\right)}{3 (1-x)^4}
H_{1,1,0,0} \nonumber \\ &\quad
+\frac{8 x \left(1+12 x+x^2\right)}{3 (1-x)^4}
H_{1,0,-1,0}+\frac{2 x \left(5-14 x+5 x^2\right)}{(1-x)^4}
H_{1,0,1,0}+\frac{4 x \left(11-6 x+11 x^2\right)}{3 (1-x)^4}
 \nonumber \\ &\quad\quad
H_{1,0,0,1}+\frac{x \left(53-259 x+187 x^2-29 x^3\right)}{3 (1-x)^5}
H_{1,0,0,0}+\frac{4 x \left(1-86 x+82 x^2+25 x^3\right)}{9 (1-x)^5}
 \nonumber \\ &\quad\quad
H_{0,1,0,0}+\frac{8 x \left(25+15 x-33 x^2+3 x^3+11 x^4\right)}{9 (1-x)^6}
H_{0,0,-1,0}-\frac{2 x (1+x) \left(1-3 x^2\right)}{3 (1-x)^5}
 \nonumber \\ &\quad\quad
H_{1,0,0,0,0}-\frac{16 x (1+x) \left(5+3 x^2\right)}{3 (1-x)^5}
H_{0,0,-1,-1,0}-\frac{16 x \left(1+7 x-5 x^2+x^3\right)}{3 (1-x)^5}
H_{0,1,0,-1,0} \nonumber \\ &\quad
+\frac{16 x \left(1-4 x+x^2\right)}{3 (1-x)^4}
H_{1,0,0,0,1}+\frac{16 x (1+x) \left(1+3 x^2\right)}{3 (1-x)^5}
H_{0,0,-1,0,1} \nonumber \\ &\quad
+\frac{4 x (1+x) \left(3+4 x^2\right)}{3 (1-x)^5}
H_{0,0,1,0,0}+\frac{16 x \left(3-3 x+7 x^2+x^3\right)}{3 (1-x)^5}
H_{1,0,0,-1,0} \nonumber \\ &\quad
+\frac{4 x \left(7+x+12 x^2+6 x^3\right)}{(1-x)^5}
H_{0,1,0,0,0}+\frac{4 x \left(7-5 x+19 x^2+7 x^3\right)}{3 (1-x)^5}
H_{1,0,1,0,0} \nonumber \\ &\quad
+\frac{4 x \left(7-17 x+31 x^2+7 x^3\right)}{3 (1-x)^5}
H_{0,1,0,0,1}+\frac{4 x \left(11+5 x+17 x^2+11 x^3\right)}{3 (1-x)^5}
H_{0,1,0,1,0} \nonumber \\ &\quad
+\frac{4 x \left(13+x+25 x^2+13 x^3\right)}{3 (1-x)^5}
H_{0,1,1,0,0}+\frac{2 x (1+x)}{3 (1-x)^3}\Big[
29 H_{-1,0}+19 H_{0,1}+2 H_{1,0} \nonumber \\ &\quad\quad
+6 H_{0} L_\mu\Big]+\frac{x (1+x) \left(11-8 x+11 x^2\right)}{9 (1-x)^5}\Big[
\pi ^2 H_{-1}-12 H_{-1,-1,0}-12 H_{-1,0,1}\Big] \nonumber \\ &\quad
+\frac{4 x \left(1+4 x-2 x^2+x^3\right)}{9 (1-x)^5}\Big[
\pi ^2 H_{0,1,0}-6 H_{1,0,0,1,0}\Big]+\frac{8 x^2}{3 (1-x)^4}\Big[
\pi ^2 H_{1,1,0}-9 H_{1,1,0,0,0} \nonumber \\ &\quad\quad
-24 H_{1,1,0,-1,0}-24 H_{1,1,0,0,1}-6 H_{1,1,0,1,0}-12 H_{1,1,1,0,0}-12 H_{1,1} \zeta_{3}\Big] \nonumber \\ &\quad
+\frac{2 x (1+x)^2}{9 (1-x)^4}\Big[
2 \pi ^2 H_{0,1}+2 \pi ^2 H_{0,0,-1}+4 \pi ^2 H_{1,0,-1}+48 H_{0,-1,1,0}+48 H_{0,1,-1,0} \nonumber \\ &\quad\quad
-6 H_{0,1,0,1}-18 H_{0,1,1,0}-48 H_{1,0,-1,-1,0}-48 H_{1,0,-1,0,1}+\pi ^2 H_{0} L_\mu+24 H_{0,-1,0} L_\mu \nonumber \\ &\quad\quad
-6 H_{0,1,0} L_\mu\Big]+\frac{x (1+x) \left(1+x^2\right)}{270 (1-x)^5}\Big[
-18 \pi ^4 H_{-1}+23 \pi ^4 H_{0}-120 \pi ^2 H_{-1,0,0} \nonumber \\ &\quad\quad
+300 \pi ^2 H_{0,0,0}+240 \pi ^2 H_{0,0,1}+1440 H_{-1,0,-1,0,0}-2880 H_{-1,0,0,-1,0}+720 H_{0,0,0,0,0} \nonumber \\ &\quad\quad
-180 H_{-1,0,0,0,0}+720 H_{-1,0,0,1,0}-2520 H_{-1,0,1,0,0}+1440 H_{0,-1,-1,0,0}+18 \pi ^4 L_\mu \nonumber \\ &\quad\quad
+1440 H_{0,-1,0,-1,0}-3960 H_{0,-1,0,0,0}-1440 H_{0,-1,0,0,1}-2160 H_{0,-1,0,1,0} \nonumber \\ &\quad\quad
-1440 H_{0,-1,1,0,0}+4320 H_{0,0,-1,0,0}+5760 H_{0,0,-1,1,0}+7200 H_{0,0,0,-1,0} \nonumber \\ &\quad\quad
+180 H_{0,0,0,0,1}-1620 H_{0,0,0,1,0}+5760 H_{0,0,1,-1,0}-720 H_{0,0,1,0,1}-2160 H_{0,0,1,1,0} \nonumber \\ &\quad\quad
-1440 H_{0,1,-1,0,0}-1440 H_{1,0,-1,0,0}+120 \pi ^2 H_{0,0} L_\mu-1440 H_{0,-1,0,0} L_\mu \nonumber \\ &\quad\quad
+2880 H_{0,0,-1,0} L_\mu+180 H_{0,0,0,0} L_\mu-720 H_{0,0,1,0} L_\mu+2520 H_{0,1,0,0} L_\mu \nonumber \\ &\quad\quad
-1440 H_{-1,0} \zeta_{3}-360 H_{0,-1} \zeta_{3}+2520 H_{0,0} \zeta_{3}+1440 H_{0} L_\mu \zeta_{3}\Big]
 \label{eq:Cggh21cF} \,,
\end{align}}%
where the quark mass has been renormalized in the on-shell scheme.
The contribution proportional to the color factor $C_A$ is given by
{\allowdisplaybreaks\begin{align}
\mathcal{C}_{\ggh}^{(C_A)} &= 
-\frac{4 x (3+x) (1+3 x)}{3 (1-x)^4}
L_\mu \zeta_{3}-\frac{x \left(79+193 x-202 x^2-112 x^3\right)}{9 (1-x)^5}
\zeta_{3} \nonumber \\ &\quad
-\frac{x \left(283-708 x+460 x^2-420 x^3+427 x^4\right)}{3240 (1-x)^6}
\pi ^4+\frac{(17-77 x) x}{216 (1-x)^3}
\pi ^2 \nonumber \\ &\quad
-\frac{x \left(1-15 x+12 x^2-12 x^3\right)}{18 (1-x)^5}
\pi ^2 H_{0}-\frac{x \left(25-8 x-84 x^2+16 x^3+37 x^4\right)}{3 (1-x)^6}
H_{0} \zeta_{3} \nonumber \\ &\quad
+\frac{(34-41 x) x}{6 (1-x)^3}
H_{0}-\frac{x \left(25+38 x+25 x^2\right)}{27 (1-x)^4}
\pi ^2 H_{1,0} \nonumber \\ &\quad
-\frac{x \left(127-96 x-686 x^2+96 x^3+223 x^4\right)}{432 (1-x)^6}
\pi ^2 H_{0,0}+\frac{(85-91 x) x}{9 (1-x)^3}
H_{1,0} \nonumber \\ &\quad
+\frac{(97-79 x) x}{9 (1-x)^3}
H_{0,1}+\frac{x (1+x)}{(1-x)^3}
H_{-1,0}+\frac{x \left(1145-4208 x+1253 x^2\right)}{324 (1-x)^4}
H_{0,0} \nonumber \\ &\quad
-\frac{2 x \left(5+9 x+5 x^2\right)}{3 (1-x)^4}
H_{0,1,0}-\frac{2 x \left(4+18 x-15 x^2+7 x^3\right)}{3 (1-x)^5}
H_{0,0,1} \nonumber \\ &\quad
-\frac{x \left(72+248 x+283 x^2+171 x^3\right)}{18 (1-x)^5}
H_{0,0,0}+\frac{x \left(219+362 x+219 x^2\right)}{18 (1-x)^4}
H_{1,0,0} \nonumber \\ &\quad
+\frac{2 x \left(1-13 x+10 x^2-12 x^3\right)}{3 (1-x)^5}
H_{0,-1,0}-\frac{x \left(23+50 x+23 x^2\right)}{3 (1-x)^4}
H_{0,1,0,0} \nonumber \\ &\quad
-\frac{4 x \left(25+86 x+25 x^2\right)}{9 (1-x)^4}
H_{1,0,-1,0}-\frac{4 x \left(11+x-8 x^2-5 x^3+8 x^4\right)}{3 (1-x)^6}
H_{0,0,0,1} \nonumber \\ &\quad
-\frac{x \left(137+36 x-7 x^2-204 x^3+17 x^4\right)}{9 (1-x)^6}
H_{0,0,0,0} \nonumber \\ &\quad
-\frac{4 x \left(5-12 x-37 x^2+36 x^3+29 x^4\right)}{9 (1-x)^6}
H_{0,0,-1,0}+\frac{2 x \left(19+26 x+19 x^2\right)}{9 (1-x)^4}
H_{1,1,0,0} \nonumber \\ &\quad
+\frac{x \left(35+107 x-215 x^2-71 x^3\right)}{9 (1-x)^5}
H_{1,0,0,0}+\frac{x}{162 (1-x)^2}\Big[
1501+2025 H_{1}+918 L_\mu \nonumber \\ &\quad\quad
+3168 H_{1,1}+792 H_{0} L_\mu+1584 H_{1} L_\mu+396 L_\mu^{2}\Big]+\frac{8 x \left(1-4 x+x^2\right)}{9 (1-x)^4}\Big[
-2 H_{1,0,1,0} \nonumber \\ &\quad\quad
-H_{1,0,0} L_\mu\Big]+\frac{x (1+x) \left(11-8 x+11 x^2\right)}{18 (1-x)^5}\Big[
-\pi ^2 H_{-1}+12 H_{-1,-1,0}+12 H_{-1,0,1}\Big] \nonumber \\ &\quad
+\frac{2 x \left(11+13 x+5 x^2-5 x^3\right)}{3 (1-x)^5}\Big[
-2 H_{0,0,1,0}-H_{0,0,0} L_\mu\Big] \nonumber \\ &\quad
+\frac{x \left(3-4 x+16 x^2-4 x^3+3 x^4\right)}{9 (1-x)^6}\Big[
-\pi ^2 H_{0,-1}+12 H_{0,-1,-1,0}+12 H_{0,-1,0,1}\Big] \nonumber \\ &\quad
+\frac{x (1+x)^2}{540 (1-x)^4}\Big[
-43 \pi ^4 H_{0}-110 \pi ^4 H_{1}-120 \pi ^2 H_{0,0,-1}-270 \pi ^2 H_{0,0,0}-240 \pi ^2 H_{0,0,1} \nonumber \\ &\quad\quad
-600 \pi ^2 H_{0,1,0}-240 \pi ^2 H_{1,0,-1}-720 \pi ^2 H_{1,0,0}-480 \pi ^2 H_{1,0,1}-1200 \pi ^2 H_{1,1,0} \nonumber \\ &\quad\quad
-5280 H_{0,0,1,1}-5280 H_{0,1,0,1}-5280 H_{0,1,1,0}-960 H_{1,0,0,1}+4320 H_{0,0,-1,-1,0} \nonumber \\ &\quad\quad
-3600 H_{0,0,-1,0,0}-2880 H_{0,0,-1,1,0}-4320 H_{0,0,0,-1,0}-1440 H_{0,0,0,0,0}-32 \pi ^4 L_\mu \nonumber \\ &\quad\quad
-1080 H_{0,0,0,0,1}-540 H_{0,0,0,1,0}-2880 H_{0,0,1,-1,0}+1620 H_{0,0,1,0,0}-7200 H_{0,1,0,-1,0} \nonumber \\ &\quad\quad
+2520 H_{0,1,0,0,0}+8640 H_{1,0,-1,-1,0}-7200 H_{1,0,-1,0,0}-5760 H_{1,0,-1,1,0}-60 \pi ^2 \zeta_{3} \nonumber \\ &\quad\quad
-14400 H_{1,0,0,-1,0}+3960 H_{1,0,0,0,0}+2880 H_{1,0,0,1,0}-5760 H_{1,0,1,-1,0}-810 H_{0,0} L_\mu \nonumber \\ &\quad\quad
+2880 H_{1,0,1,0,0}-14400 H_{1,1,0,-1,0}+7200 H_{1,1,0,0,0}-120 \pi ^2 H_{0,0} L_\mu-660 H_{0,0} L_\mu^{2} \nonumber \\ &\quad\quad
-240 \pi ^2 H_{1,0} L_\mu-2640 H_{0,0,1} L_\mu-2640 H_{0,1,0} L_\mu-1440 H_{0,0,-1,0} L_\mu-12240 H_{1} \zeta_{3} \nonumber \\ &\quad\quad
-360 H_{0,0,0,0} L_\mu-2880 H_{1,0,-1,0} L_\mu+1440 H_{1,0,0,0} L_\mu-4875 H_{0,0} \zeta_{3}-10800 H_{0,1} \zeta_{3} \nonumber \\ &\quad\quad
-10800 H_{1,0} \zeta_{3}-21600 H_{1,1} \zeta_{3}-2160 H_{0} L_\mu \zeta_{3}-4320 H_{1} L_\mu \zeta_{3}+360 \zeta_{5}\Big]
 \label{eq:Cggh21cA}
  \,.
\end{align}}%
The results with an \MSbar renormalized quark mass is
\begin{align}
  \overline{\mathcal{C}}_{\ggh}^{(C_F)}
  &=
  \mathcal{C}_{\ggh}^{(C_F)}
  +
  \Delta \mathcal{C}_{\ggh}^{(C_F)}
  \,, \\
  \overline{\mathcal{C}}_{\ggh}^{(C_A)}
  &=
  \mathcal{C}_{\ggh}^{(C_A)}
  \,,
\end{align}
where
\begin{align}
\Delta \mathcal{C}_{\ggh}^{(C_F)} &= 
-\frac{x (3+11 x)}{2 (1-x)^3}
H_{0} L_\mu-\frac{x (7+135 x)}{24 (1-x)^3}
H_{0}-\frac{x \left(11+42 x+3 x^2\right)}{2 (1-x)^4}
H_{0,0} L_\mu \nonumber \\ &\quad
-\frac{x \left(135+426 x+7 x^2\right)}{24 (1-x)^4}
H_{0,0}+\frac{x}{12 (1-x)^2}\Big[
71+8 \pi ^2+64 H_{1}+84 L_\mu+48 H_{1} L_\mu \nonumber \\ &\quad\quad
+36 L_\mu^{2}\Big]+\frac{x (1+x)}{6 (1-x)^3}\Big[
-2 \pi ^2 H_{0}-16 H_{0,1}-16 H_{1,0}-12 H_{0,1} L_\mu-12 H_{1,0} L_\mu \nonumber \\ &\quad\quad
-9 H_{0} L_\mu^{2}\Big]+\frac{x \left(1+6 x+x^2\right)}{6 (1-x)^4}\Big[
-2 \pi ^2 H_{0,0}-24 H_{0,0,0}-16 H_{0,0,1}-16 H_{0,1,0} \nonumber \\ &\quad\quad
-16 H_{1,0,0}-18 H_{0,0,0} L_\mu-12 H_{0,0,1} L_\mu-12 H_{0,1,0} L_\mu-12 H_{1,0,0} L_\mu-9 H_{0,0} L_\mu^{2}\Big]

  \,.
\end{align}
The contribution $\mathcal{C}_{\ggh}^{(C_A)}$ is renormalization scheme independent.

For the sake of completeness, we also provide the $n_l^2$-contribution
$\tilde C_{\ggh}^{(2,2)}$ originating from the infrared subtraction,
cf.\ eq.~(\ref{eq:cgghnlsplit}):
\begin{align}
  \tilde C_{\ggh}^{(2,2)}
  &=
  T_F^3\Big\{
\frac{x}{54 (1-x)^2}\Big[
-\pi ^2-48 H_{0,0}-96 H_{0,1}-96 H_{1,0}-192 H_{1,1}-48 H_{0} L_\mu-24 L_\mu^{2} \nonumber \\ &\hspace{2.5em}\quad
-96 H_{1} L_\mu\Big]+\frac{x (1+x)^2}{108 (1-x)^4}\Big[
\pi ^2 H_{0,0}+288 H_{0,0,0,0}+288 H_{0,0,0,1}+288 H_{0,0,1,0} \nonumber \\ &\hspace{2.5em}\quad
+192 H_{0,0,1,1}+288 H_{0,1,0,0}+192 H_{0,1,0,1}+192 H_{0,1,1,0}+288 H_{1,0,0,0} \nonumber \\ &\hspace{2.5em}\quad
+192 H_{1,0,0,1}+192 H_{1,0,1,0}+192 H_{1,1,0,0}+144 H_{0,0,0} L_\mu+96 H_{0,0,1} L_\mu \nonumber \\ &\hspace{2.5em}\quad
+96 H_{0,1,0} L_\mu+96 H_{1,0,0} L_\mu+24 H_{0,0} L_\mu^{2}\Big]

  \Big\}
  \,.
\end{align}


\section{Ancillary File} \label{sect:appdx:anc}
The ancillary file \texttt{ggh-aah-nl.m} contains the main results of
this paper in an electronic form readable
by \texttt{Mathematica}.\footnote{Wolfram Research, Inc., \emph{Mathematica, Version 12.0}, Champaign, IL, U.S.A.} The following table
describes its notation:
\begin{center}
  \begin{longtable}{@{}p{10cm}p{3cm}@{}}
    \toprule
    \texttt{CaahOS} (\texttt{CaahMSbar}) & $C_{\aah}$ \\
    \texttt{CaahOS0} (\texttt{CaahMSbar0}) & $C^{(0)}_{\aah}$ \\
    \texttt{CaahOS1} (\texttt{CaahMSbar1}) & $C^{(1)}_{\aah}$ \\
    \texttt{CaahOS2} (\texttt{CaahMSbar2}) & $C^{(2)}_{\aah}$ \\
    \texttt{caahOS2nl1nonsing} (\texttt{caahMSbar2nl1nonsing}) & $\mathcal{C}_{\aah}^{(\text{non-sing})}$ \\
    \texttt{caahOS2nl1sing} (\texttt{caahMSbar2nl1sing}) & $\mathcal{C}_{\aah}^{(\mathrm{sing})}$ \\ \midrule
    \texttt{CgghOS} (\texttt{CgghMSbar}) & $\tilde C_{\ggh}$ \\
    \texttt{CgghOS0} (\texttt{CgghMSbar0}) & $\tilde C^{(0)}_{\ggh}$ \\
    \texttt{CgghOS1} (\texttt{CgghMSbar1}) & $\tilde C^{(1)}_{\ggh}$ \\
    \texttt{CgghOS2} (\texttt{CgghMSbar2}) & $\tilde C^{(2)}_{\ggh}$ \\
    \texttt{cgghOS2nl1cF} (\texttt{cgghMSbar2nl1cF}) & $\mathcal{C}_{\ggh}^{(C_F)}$ \\
    \texttt{cgghOS2nl1cA} (\texttt{cgghMSbar2nl1cA}) & $\mathcal{C}_{\ggh}^{(C_A)}$ \\ \midrule
    \texttt{api} & $\alpha/\pi$ \\
    \texttt{aspi} & $\alpha_s/\pi$ \\
    \texttt{Lmu} & $L_\mu$ \\
    \texttt{QQ} & $Q_{\mathrm{q}}^2$ \\
    \texttt{QQ2} & $\sum_{i=1}^{n_l} Q_i^2$ \\
    \texttt{HPL[a\_List,x]} & $H_{\vec a}$ \\
    \bottomrule
  \end{longtable}
\end{center}
The \texttt{OS}\,(\texttt{MSbar}) versions correspond to a quark mass renormalized in the on-shell\,(\MSbar) scheme.
The harmonic polylogarithms are in a format compatible with the \texttt{Mathematica} package \texttt{HPL.m}~\cite{Maitre:2005uu}.



\bibliography{biblio}

\end{document}